# Photons Squeezing Model in a Nonlinear Micro-ring Resonator

W. Khunnam,[a] J. Ali,[b, c] and P. Yupapin [d, e]*

[a]Department of Physics, Faculty of Science, Naresuan University, Pitsanulok, Thailand;
[b]Laser Centre, IBNU SINA ISIR, Universiti Teknologi Malaysia 81310 Johor Bahru, Malaysia
[c]Faculty of Science, Universiti Teknologi Malaysia, 81310 Johor Bahru, Malaysia;
[d]Department for Management of Science and Technology Development, Ton Duc Thang University, District 7, Ho Chi Minh City, Vietnam;
[e]Faculty of Electrical & Electronics Engineering, Ton Duc Thang University, District 7, Ho Chi Minh City, Vietnam;
*Corresponding author E-mail: <preecha.yupapin@tdt.edu.vn>

**Abstract:** The use of a modified add-drop filter configured by the coupled two side rings for photon squeezing system is proposed. There are two forms of the coupling energies when photons travel around the micro-ring resonator that it is coupled by the two nonlinear side rings. They are the external disturbances known as the construction(creation, $\hat{a}^{\dagger}$) and destruction (annihilation, $\hat{a}$) energy operators, which can disturb the traveling photons on the nonlinear microring resonator, from which the nonlinear behaviour called a four-wave mixing (FWM) is introduced by the coupled rings, which leads the system being unstable from the harmonic motion, from which the squeezed photons can occur. In a simulation, the obtained result of such a proposed behavior is confirmed by using the commercial Opti-wave and MATLAB software programs, whereas the suitable simulation parameters were chosen, the quantum harmonic squeezed photon theory is also reviewed, and the numerical details were given.

## 1. Introduction

In an optical system, there is a problem called the quantum noise always occurred, which is introduced by the photon traveling in the system with the oscillation angular frequency, $\omega$. The squeezed photons can be generated within the system whenever the uncertainty value ($\Delta x. \Delta p$) is saturated [1, 2], which is formed by the nonlinear effect known as a four-wave mixing (FWM). Theoretically, the quantum noise is introduced by squeezed photons and affected to the optical system performance, therefore all optical systems are required to operate over the quantum noises, which is the required optical system specification. On the other hands, this is the system limitation that is required to be solved. When a light pulse from a laser source in either a common (Gaussian pulse) or fiber laser (soliton pulse) is input into the micro-ring resonator (waveguide) system, such a light pulse behavior can be described and interpreted by using well-established mathematical model [3-6], from which the dynamic of a pulse within the device and output of light via each port can be found by the different interpretations. Firstly, the transfer function of the output can be obtained by using the Meson's rule [7], which is the well-known ray tracing technique. Secondly, the wave propagation within the device and the output light can be described by using the wave equation, where the transverse electric and magnetic fields can be obtained. Moreover, the leaky mode of the waves within the ring system can also be known and the optimum leaky mode known as the whispering gallery mode (WGM) can also be obtained by controlling the specific conditions [3, 5 7]. Finally, the wave-particle behaviors of light pulse can also be interpreted by using in either the Schrodinger equation or Dirac notation approach [8, 9], whereas the interesting aspect is that when the tunneling photons or leaky modes of light are collected within the center ring, which is called the whispering gallery mode(WGM) of light(or squeezed photons). In this article, the main point is that the tunneling photons can be squeezed from the nonlinear ring system, whereas the nonlinear effect is introduced by the coupled two side rings, wherein this proposal the *GaAsInP/P* material is proposed for the nonlinear material device. The coupling effect is introduced by the two side rings and into the master ring, from which the balance of the effect is formed by the common path cancellation and presented by the annihilation ($\hat{a}$) and the creation ($\hat{a}^{\dagger}$) operation concept in the quantum harmonic squeezing system [10, 11]. However, the coupling effect between the two inside rings can be avoided by given the reasonable gap of between them, which is the important issue being taken care. There are the other states of squeezed photon sates called the excited squeezed states by the FWM effects introduced by the creation and annihilation enforcements [1, 2, 10, 11] which will introduce the changes in frequency oscillations (energy of state). However, in this article the vacuum squeezed photon state is introduced for a simple investigation, while the higher photon squeezed states will be our continuing research work.

## 2. Analytical Model

A light pulse in the form of the slow varying amplitude can be configured to have the wave-particle duality behaviors, which is described by the nonlinear Schrodinger equation (NLSE) [1, 2]. Theoretically, when the light pulse travels within the microring resonator, it is behaved as a quantum harmonic oscillation, in which the wave packet with slow varying amplitude is confined within the system, with the angular frequency oscillation is $\omega$. Principally, the squeezed photon state within the optical system is formed by the nonlinear phenomenon known as a four-wave mixing, where in this case, the side peaks of the center frequency are introduced by the coupled side rings that can behave the as the construction (creation) and destruction (annihilation) energies within the system, from which the superposition of all peaks can lead to build the squeezed photons, which is the external disturbance.

When an optical pulse is entered into the microring system as shown in Figure 1, the wave-particle duality is satisfied if the pulse is propagated by slow varying amplitude condition, which is described by the nonlinear Schrodinger equation (NLSE). The varying function amplitude is represented by the optical pulse (photon) [1, 2], is given by

$$\Psi(x) = C\exp\left(-\frac{(x-x_0)^2}{2\omega_0^2} + ip_0 x\right) \quad (1)$$

where C, $x_0$, $w_0$, $p_0$ are constants, namely a normalized center wave-packet, pulse width, momentum expectation, respectively. In this case the constant ℏ is one, where $\hbar = \frac{h}{2\pi}$, h is the Plank's constant. The squeezed state of the particles (photons) within the ring resonator is given by $\hat{x} + i\,\hat{p}\,w_0^2$, where the eigenvalue is $x + ip_0$.

From Figure 1, the squeezed state for a quantum harmonic oscillator of the particle within the ring resonator system with the two side rings disturbance is given by

$$|\propto, Š\rangle = D(\propto)\, S(Š)\, |0\rangle \quad (2)$$

Where $|0\rangle$ is the vacuum state, D($\propto$) is the displacement operator and S(Š) is the squeezed operator, which given by fo by $D(\propto) = \exp(\propto \hat{a}^\dagger - \propto^* \hat{a})$ and $S(Š) = \exp(\frac{1}{2}(Š^*\hat{a}^2 - Š\hat{a}^{\dagger 2}))$.

Where $\hat{a}$ and $\hat{a}^\dagger$ are the annihilation and creation operator, respectively. For a quantum operators, are given by $\hat{a}^\dagger = \sqrt{\frac{m\omega}{2\hbar}}(x - \frac{ip}{m\omega})$ and $\hat{a} = \sqrt{\frac{m\omega}{2\hbar}}(x + \frac{ip}{m\omega})$. For a real Š, the uncertainty in x and p are given by $(\Delta x)^2 = \frac{\hbar}{2m\omega}e^{-2Š}$ and $(\Delta p)^2 = \frac{m\hbar\omega}{2}e^{2Š}$. The squeezed coherent states by the "Heisenberg Uncertainty Principle" $\Delta x \Delta p = \hbar/2$, with reduced uncertainty in one of its quadrature components and increased uncertainty in the other.

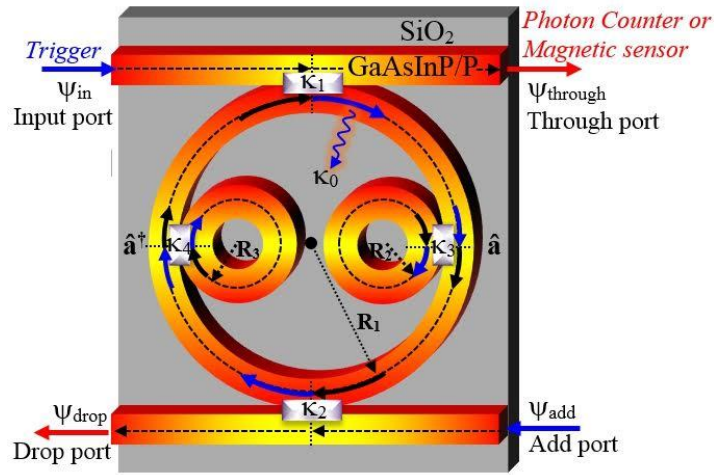

**Figure 1**: Microring structure of the proposed system, where , where R: Ring radii, κ :Coupling constant, R: Ring radius, $\hat{a}$ : annihilation operator, $\hat{a}^\dagger$: Creation operator, and Ψ : Wave function, where the wave function wave number ($κ_0$) is oscillated in the vacuum state.

From Figure 1, the quantum harmonic oscillator of angular frequency, $\omega$, which will be disturbed by the two side ring photons and coupled to the photon distribution, which can be ramdomly coupled by the creation and annihilation operations from the induced FWM behavior, when the uncertainty value is saturated [10, 11], the squeezed states are observed, from which the photons will be squeezed from the ring resonator system. The squeezed state of photons in the vacuum state is obtained by the photon oscillation in the ground state, where the squeezed photon state position is observed by oscillation along the displacement(x). Generally, there are the other states of squeezed photon sates called the excited squeezed states by the FWM effects introduced by the creation and annihilation enforcements, which will introduce the changes in frequency oscillations (energy of state), which is as shown in the below details. The magnitude of the complex term is the unity, which is not affected to the system, finally the relation $\Delta x \Delta p = \hbar/2$ is obtained, from which $\Delta x$ is approximated to be $(\hbar/2m\omega)^{1/2}$.

## 3. Simulation Results

In a simulation, a wave function in the form of the slow varying amplitude is given by Equation (1), which is input into the system as shown in Figure 1. The slow varying amplitude term is the first term of the exponent argument, while the second term is the phase term, which is the unity in magnitude. The given normalized constant is 5, the oscillation center is at $x_0$, the displacement is from x equal to 0 to 10 μm, $\omega$ is $2\pi\nu$, which is equal to $2\pi c/\lambda$, where ν is the linear oscillation frequency, c is the speed of light in vacuum, λ is the light source wavelength, which is 1.55 μm. The plot between the photon distribution and oscillation displacement is as shown in Figure 3, which is obtained by the relationships between the photon distribution ($|\Psi|^2$), which is $5\exp-\frac{(x-x_0)^2}{2\omega_0^2}$, the displacement is obtained from the uncertainty terms, while the decayed terms is become the unity.

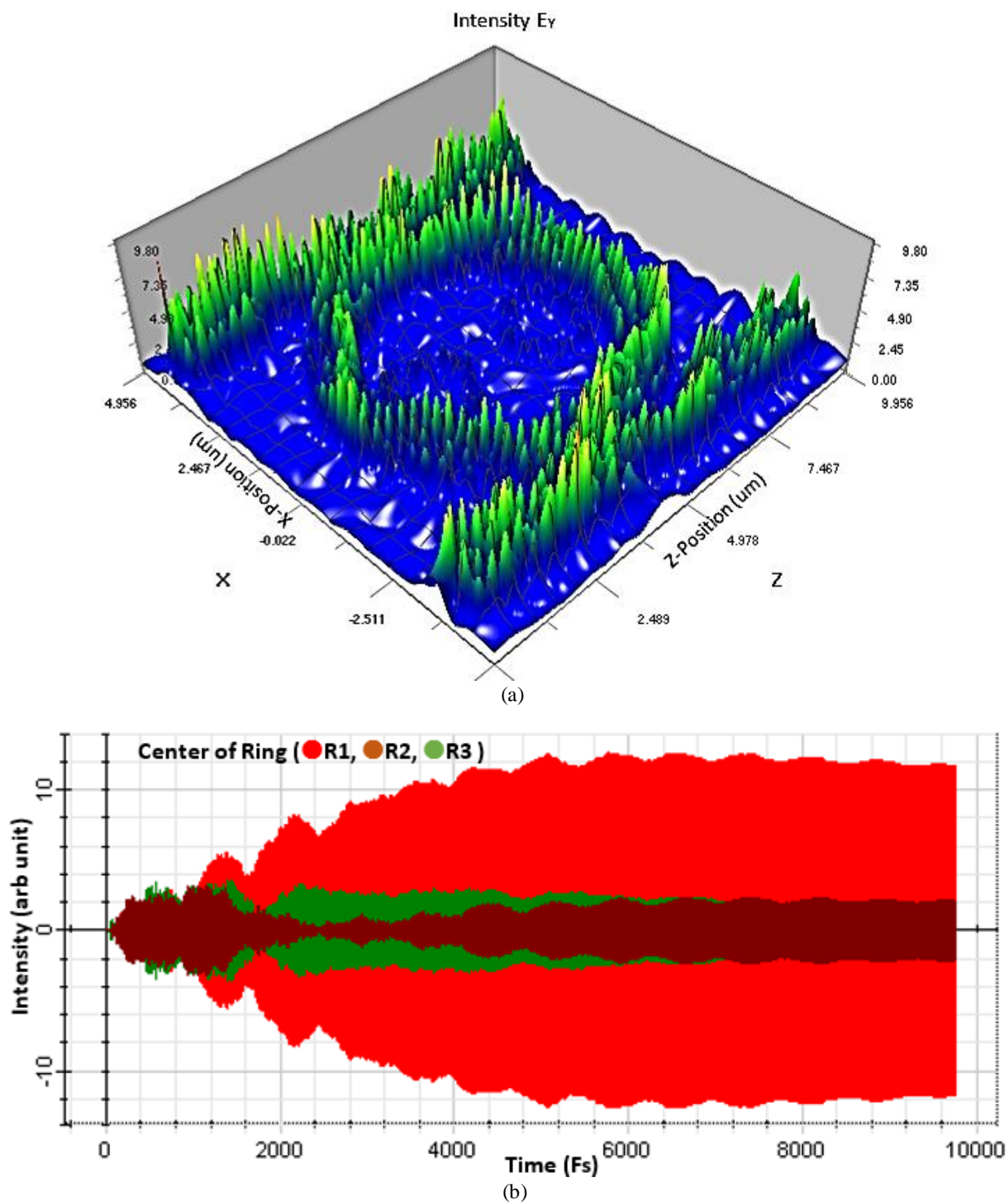

(a)

(b)

**Figure 2:** The Opti-wave simulation result of the system in Figure 1 and squeezing photons in time domain

Such a phenomenon can be confirmed by the plot of the relationship between the photon distribution and oscillation distance, which is terms of the uncertainty [1, 2], which is given in the next section. Generally, the two coupling nonlinear rings can be located in either outside or inside the master ring, where the use of outside coupling resonator has reported for many applications [12-15], from which the whispering gallery mode of light could also be obtained by controlling the suitable parameters [12], especially, when the two outside ring parameters such as coupling coefficients and ring radii were matched. In Figures 2 and 3, show the squeezing photon from the micro-ring system with respect to time and wavelength, respectively, where the squeezed photons have seen in $R_1$ and $R_3$, there is no squeezed photon in $R_3$. In Figure 4, the obtained results have shown that the squeezed state is obtained when x are 3, 8 and -4.5, 0, 4.5 μm on the Z and X axes of the system in Fig. 1(a), respectively. In practice, the change in the other parameters can be arranged, the other squeezed positions will be observed, which can be useful for different application aspects. The squeezed state of photons in the vacuum state is obtained by the photon oscillation in the ground state, which is described by an Equation (1), where the squeezed photon state position is observed by oscillation along the displacement(x).

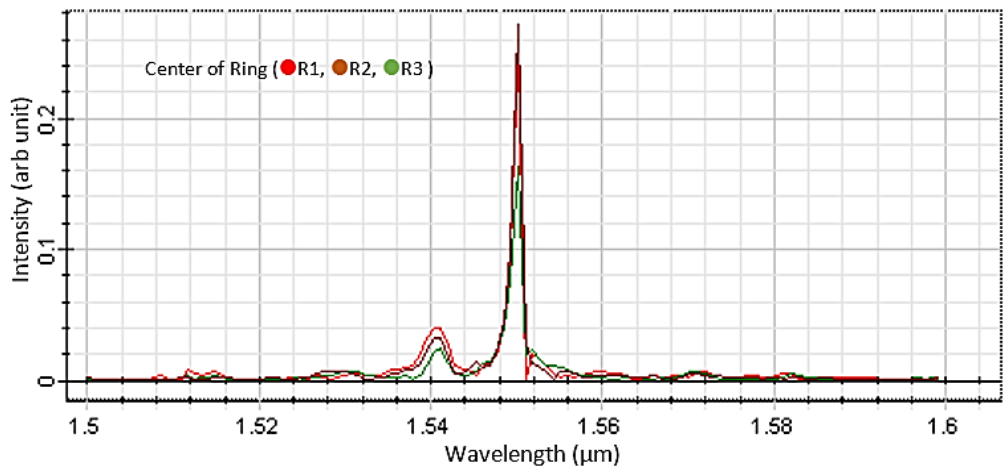

**Figure 3**: Plot of the relationship between photon squeezing and wavelength

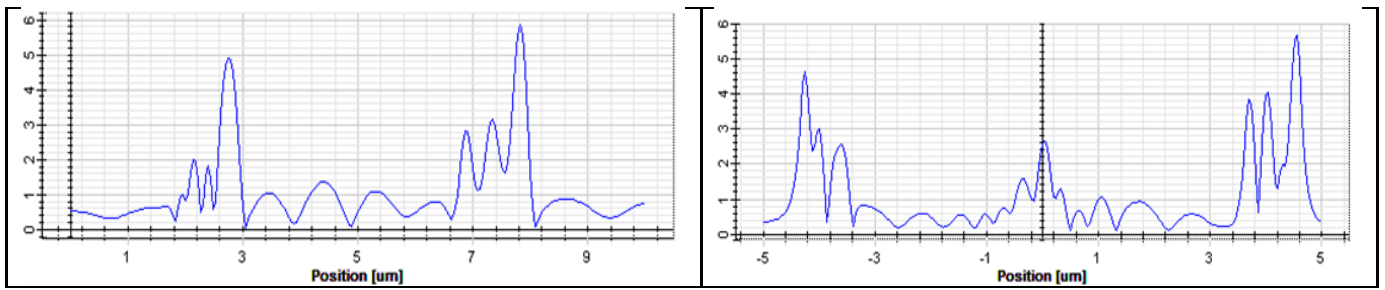

**Figure 4**: Plot between photon squeezing displacement (ring diameter)

### 4. Discussion and Conclusion
In this model, the inside coupling ring scheme is proposed, in which the coupling effect between two inside rings is neglected because when the case the wave-particle duality behavior is applied, in which the photon is now presented within the ring system, which is much smaller than the propagation wavelength. Principally, the nonlinear effect is introduced by the two side rings and coupled into the master ring, which can occur within the gap between the coupling rings. In this structure, the construction and destruction energies are substituted by the two different operators in the quantum harmonic oscillator, which can lead the quantum squeezing occurrence, from which the tunneling photon is induced by the photon and the energy is greater than the system potential energy [1, 2]. For further applications, the two side rings in the inside coupling scheme are required to give the photons within the center ring being excited, wherein the nonlinear material such as *GaAsInP/P* is given, which can be also used as the transparent material for light propagation, while the other material is a graphene material, is the very good candidate for the new era of electronics. However, to avoid the coupling effect between the collected electrons and coupled ring material, the two side rings are suggested to use the other nonlinear materials. In operation, the tunneling photons are collected within the center ring and the beam of energy formed, where they are four different squeezing forms, which are (i) the vacuum state squeezing, $|0>$, which is the ground state squeezing and rarely occurred, (ii) the amplitude squeezing is the resonance in signal amplitudes within the system,

(iii) the phase squeezing is the phase resonance occurred, and finally (iv) the arbitrary squeezing is the randomly squeezed by the resonance in either phase or amplitude signals. The obtained outputs of the system as shown in Figure 1 are categorized into two forms, which are (i) electrons and their spins and (ii) photons, can be observed by using the magnetic sensor and photon counter at the output ports, respectively, moreover the collection of electrons or photons at the center ring are also available for applications such as electron source, photon source, and energy storage. In the quantum scheme, the trigger is required for the initial input photon into the system to synchronize between the initial input photon and the receiver, where the quantum entangled codes can be recognized. The photon outputs are presented in terms of the probability of photon detection, which is given by $<\Psi|\Psi^*>$ [8], while the electron spin is detected by the magnetic sensor.

The use of a microring resonator is proposed for a photon squeezing system, from which a system is formed by a nonlinear micro-ring resonator, which can be fabricated by present technology in the form of stacked waveguides, the dimension is few hundreds micrometer. By using the commercial software programs with the suitable selected parameters, the simulation results are obtained, which is confirmed that the photon squeezing can be generated based on the nonlinear behavior called a FWM within the system, which can be useful for further applications, where the use for quantum computer and information may be realized.

**Acknowledgment**
The authors would like to give the acknowledgment to the Ton Duc Thang University, Ho Chi Minh City, Vietnam for the research facilities.